\documentclass[fleqn,twoside]{article}
\usepackage{espcrc2}
\usepackage{graphicx}
\usepackage{epsfig}
\def\etal{\hbox{\it et al.}} 

\def\epj#1#2#3{    {\it Eur. Phys. J. }{\bf #1} (#2) #3} 
\def\np#1#2#3{    {\it Nucl. Phys. }{\bf #1} (#2) #3} 
 
\def\pl#1#2#3{    {\it Phys. Lett. }{\bf #1} (19#2) #3} 
 
\def\pr#1#2#3{    {\it Phys. Rev. }{\bf #1} (#2) #3} 
 
\def\prl#1#2#3{    {\it Phys. Rev. Lett. }{\bf #1} (19#2) #3}

\newcommand{\com}[1]{ \par }

\def\evg{\, g_{eV}^\gamma} 
\def\tvg{\, g_{\tau V}^\gamma} 
\def\evz{\, g_{eV}^Z} 
\def\eaz{\, g_{eA}^Z} 
\def\tvz{\, g_{\tau V}^Z} 
\def\taz{\, g_{\tau A}^Z} 
\def\tz{\, g_{\tau T}^Z} 
\def\tg{\, g_{\tau T}^\gamma} 
\def\evg2{(g_{eV}^\gamma)^2} 
\def\tvg2{(g_{\tau V}^\gamma)^2} 
\def\evz2{(g_{eV}^Z)^2} 
\def\eaz2{(g_{eA}^Z)^2} 
\def\tvz2{(g_{\tau V}^Z)^2} 
\def\taz2{(g_{\tau A}^Z)^2} 
\def\tz2{(g_{\tau T}^Z)^2} 
\def\tg2{(g_{\tau T}^\gamma)^2}

\newcommand{\beq}{\begin{equation}} 
\newcommand{\eeq}{\end{equation}} 
\newcommand{\bi}{\begin{itemize}} 
\newcommand{\ei}{\end{itemize}} 
\newcommand{\bea}{\begin{eqnarray}} 
\newcommand{\eea}{\end{eqnarray}} 
\newcommand{\bes}{\begin{eqnarray*}} 
\newcommand{\ees}{\end{eqnarray*}}   
\def\etal{\hbox{\it et al.}} 
 
\title{$\tau$ EDM at Low Energies} 
\author{J. Bernab\'eu\address[Valencia]{Departament de F\'{\i}sica Te\`orica,  
CSIC-Universitat de Val\`encia,\\ E-46100 Burjassot, Val\`encia, Spain
\\ and\\ IFIC, centre Mixt Universitat de Val\`encia-CSIC, Valencia, Spain }, 
Gabriel A. Gonz\'alez-Sprinberg\address{ Instituto de F\'{\i}sica, Facultad de Ciencias, 
Universidad de la Rep\'ublica, \\ 
Igu\'a 4225, 11400 Montevideo, Uruguay } and
Jordi Vidal\addressmark[Valencia]}
\begin{document}
\begin{abstract} 
Low energy tau pair production, at B factories and on top of the
 $\Upsilon$ resonances,
 allows for a detailed investigation on
  the CP violation at the electromagnetic
 tau pair production vertex. High statistic available 
 at low energies offers the opportunity for an
 independent analysis of CP-violation in the $\tau$ lepton physics. 
We show that stringent and independent bounds on the $\tau$ electric
 dipole moment, competitive with the high energy measurements, 
can be  established in low energies experiments.
\vspace{1pc}\end{abstract} 
\maketitle
The electric dipole moment (EDM)  has been
extensively investigated in particle  physics
\cite{pdg,{Khriplovich:ga},bern}. 
Nowadays the most precise bound is the
one on the electron EDM,  $d^e_\gamma  = (0.07\pm0.07) \times
10^{-26}\,e\,cm$,
 while the loosest is on the $\tau$ EDM \cite{pdg}, 
$Re (d^\tau_\gamma) > -3.1$ and $< 3.1 \times 10^{-16}\,e\,cm$. The PDG
quoted bound on $ d^e_\gamma$ for the $\tau$ comes from a CP-even observable: 
the total cross section  for $e^+e^- \rightarrow \tau^+\tau^-\gamma$. 
This should be superseded with  the measurements of CP-odd observables, and
this is what we propose in what follows. Low energy experiments provide
 for independent constraints on the EDM and allow to separate the
effects coming from the electric and weak-electric dipole moments. 
A non zero measurement of an EDM is a time  reversal odd signal; CPT
theorem for  quantum field theories states  that this is equivalent to
CP violation.  These dipole moments are  generated in the standard model
only at three loops but extended models  can induce an  EDM not far
from present experimental sensitivities\cite{dumm}. 
The tau EDM and weak-EDM have been studied in detail at LEP in spin 
correlation observables  \cite{heidel} and also 
in spin linear terms \cite{nt}. Most of the statistics 
for the tau pair production was dominated by LEP but nowadays 
the situation has evolved. High luminosity B factories and 
their upgrades have a large $\tau$ pair sample.


We parametrized deviations from the
standard model, at low energies,  by  an effective Lagrangian built with
the standard model
particle spectrum, having as zero order term just the standard model
Lagrangian, and containing higher dimension gauge invariant operators  
suppressed by the scale  of new physics,  $\Lambda$ \cite{efflag,arcadi}. The
leading non-standard 
effects that contribute to the EDM and  weak-EDM come from dimension six
operators:

\begin{eqnarray}
\label{eq:ob}
{\cal O}_B &=& \frac{g'}{2\Lambda^2} \overline{L_L} \varphi
\sigma_{\mu\nu}
 \tau_R B^{\mu\nu}  \\ {\cal O}_W &=&  \frac{g}{2\Lambda^2}
\overline{L_L}
\vec{\tau}\varphi
\sigma_{\mu\nu} \tau_R \vec{W}^{\mu\nu}  ~.
\end{eqnarray}
Here $L_L=(\nu_L,\tau_L)$ is the tau leptonic doublet, 
$\varphi$  is the Higgs doublet,  $B^{\nu\nu}$ and $\vec{W}^{\mu\nu}$
are the 
U(1)$_Y$ and SU(2)$_L$ field strength  tensors, and $g'$ and $g$ are the
gauge 
couplings. 


The effective Lagrangian is 
\begin{equation}
\label{eq:leff}
{\cal L}_{eff} = i \alpha_B {\cal O}_B  + i \alpha_W {\cal O}_W +
\mathrm{h.c.}
\label{eq:interaccio}
\end{equation}
where the couplings
$\alpha_B$ and $\alpha_W$ real. 
After spontaneous symmetry breaking, the Higgs gets a vacuum expectation
value 
$<\varphi^0>=u/\sqrt{2}$ with $u=1/\sqrt{\sqrt{2}G_F}=246$~GeV, 
and the interactions (\ref{eq:interaccio}) can be written in terms of the gauge 
boson mass eigenstates 
$A^\mu$ and $Z^\mu$.
Thus, the Lagrangian for the EDM, written in terms
of the mass eigenstates, is\footnote{Similar results, but for the 
magnetic moments, are
found in \cite{arcadi} where the notation is the same.}
\begin{eqnarray}
{\cal L}_{eff}^{\gamma, Z} &=& 
- i \frac{e}{2 m_\tau} \,F^\tau_\gamma \,\overline{\tau} \sigma_{\mu\nu}
 \gamma^5 \tau F^{\mu\nu} -\nonumber\\
 & &i \frac{e}{2 m_\tau} \,F^\tau_Z \,\overline{\tau} \sigma_{\mu\nu}
\gamma^5 \tau
Z^{\mu\nu}
\label{eq:leff_fin}
\end{eqnarray}
where $F_{\mu\nu}$  and 
$Z_{\mu\nu}=\partial_\mu Z_\nu-\partial_\nu Z_\mu$ 
are  the Abelian field strength tensor of the photon and
 the $Z$ gauge boson, respectively.
As usual, we have defined the following dimensionless 
couplings
\begin{eqnarray}
F^\tau_\gamma &=& (\alpha_B - \alpha_W) \frac{u
m_\tau}{\sqrt{2}\Lambda^2}~,\\
F^\tau_Z &=& - (\alpha_B s_W^2 + \alpha_W c_W^2) \frac{u
m_\tau}{\sqrt{2}\Lambda^2} \frac{1}{s_W c_W}
 \label{eq:epsilonw}
\end{eqnarray}
where $s_W =\sin\theta_W$ and $c_W = \cos\theta_W$ sine and cosine of the weak angle.
Notice that, in the effective Lagrangian approach, exactly the same couplings
that contribute to processes at high energies also
contribute to the electric dipole moment form factors, 
$F^{\mathrm new}(q^2)$, at $q^2=0$. The
difference $F^{\mathrm new}(q^2)-F^{\mathrm new}(0)$ only comes from
higher dimension operators whose effect is suppressed by powers of 
$q^2/\Lambda^2$,  as
long as $q^2\ll \Lambda^2$ as needed for the consistence of the
effective
Lagrangian approach. For this reason we make  no  distinction between
electric dipole moment and electric form factor.

The  electric and weak-electric dipole moment are
\begin{eqnarray}
d^\tau_\gamma &=& \frac{e}{2 m_\tau} F^\tau_\gamma ,\\ d^\tau_Z &=&
\frac{e}{2 m_\tau} F^\tau_Z
\end{eqnarray}
and are usually expressed in units of $e cm$.

\begin{figure}[htb]
\begin{center}
\epsfig{file=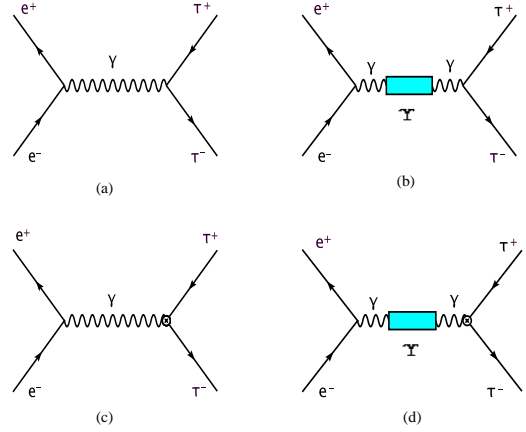,width=80mm}
\end{center}\vspace*{-1cm}
\caption{Diagrams (a) direct $\gamma$ exchange (b) $\Upsilon$
production (c) EDM in $\gamma$ exchange (d) EDM in $\Upsilon$
production
\label{fig:figura1}}
\end{figure}

The $e^+\, e^- \longrightarrow 
\gamma, \Upsilon \longrightarrow
\tau^+ \tau^-$ cross section has contributions coming from
the standard model and the effective Lagrangian 
Eq.(\ref{eq:leff_fin}). At low energies tree level contributions come from
direct  $\gamma$ exchange (off the $\Upsilon$ peak) or 
$\Upsilon$ (at the $\Upsilon$ peak) exchange while interference  
$\gamma - Z$ (or  $\Upsilon - Z$ at the $\Upsilon$ peak) and
$Z-Z$ diagrams are suppressed by $q^2/M_Z^2$. At tree level then, the
relevant  diagrams are shown in Fig.\ref{fig:figura1}. from the standard ($a$), ($b$) and
beyond the standard ($c$), ($d$)  amplitudes.

We are interested in the differential cross section for $e^+e^- 
\longrightarrow \tau^+(s_+)\tau^-(s_-)$ and we will retain only up to
linear  terms in the dipole moments. In this way,  the only proportional terms to the EDM 
one gets from diagrams $(c)$ and $(d)$ for the $e^+\, e^- \longrightarrow
\tau^+ \tau^-$ cross section come  in the spin-spin correlation. 
For details, see for example the cross section formulas in \cite{500}.

All contributions to polarization terms
are determined by the discrete symmetry properties: P, CP, T and helicity flip.
In our hypothesis the EDM does not contribute to the 
single spin dependent terms. In fact 
 it appears in the normal P-even, T-odd $(s_+-s_-)_N$ (to the scattering plane)
 spin-linear terms. Taking into accout that
  the EDM effective Lagrangian is P and T-odd, we find that
the EDM contributes to this single spin term 
 but only  through the  interference with the axial part 
of a $Z$ exchange. This 
term is  proportional to  the electron or the fermion axial 
coupling to the $Z$. As such, it is doubly suppress by $q^2/M_Z^2$ 
at low energies and also by the axial coupling ($1/4 - s_W^2$).

The EDM contributes to the spin-spin correlation terms. 
The T-odd normal-transverse $(\mbox{\boldmath $s$}_+\times \mbox{\boldmath $s$}_-)_{N,T}$ 
and normal-longitudinal $(\mbox{\boldmath $s$}_+\times \mbox{\boldmath $s$}_-)_{N,L}$ 
correlation terms are proportional to the EDM. This last term also receives
standard model contributions  through absorptive parts 
generated in radiative corrections. At tree level it is the imaginary 
part of the $Z$ propagator that produces a contribution to this correlation
with the interference of the amplitudes of direct 
$\gamma$ and $Z$ exchange. This term is suppressed at low 
energies, and has been calculated in \cite{nuria} and subtracted if necessarily.

In the following we show how to measure the EDM with the appropriate observables. 
We will follow the notation of references \cite{arcadi,nos1} and further details and results will be published elsewhere \cite{prox}. 


Let us consider the $\tau$-pair production in $e^+e^-$ collisions 
though direct $\gamma$ exchange (diagrams (a) and (b) in Fig. 1.). 

We assume from now on that the tau
production plane and direction of flight can be fully reconstructed. 
This can be done if both $\tau$ decay semileptonicaly; 
this  has been studied in \cite{kuhn} and applied
in different cases by  the L3-Collaboration \cite{l3} following the ideas of
\cite{nt,nos1} for semileptonic decays of the tau.

Polarization along the directions 
$x,y,z$ correspond to what is called transverse (T), normal (N) and longitudinal (L)
polarizations.

The differential cross section for $\tau$ pair production can be written:

\begin{equation}
\frac{d \sigma}{d  \Omega_{\tau^-}}=
\frac{d \sigma^{0}}{d  \Omega_{\tau^-}}
+\frac{d \sigma^{S}}{d  \Omega_{\tau^-}}+\frac{d \sigma^{SS}}
{d  \Omega_{\tau^-}}+\ldots 
\label{cross1}
\end{equation}
The dots symbolize higher orders in
the effective Lagrangian that are beyond experimental sensitivity and not 
considered in this paper.
The first term of Eq. (\ref{cross1}) represents the spin independent 
differential cross section
\begin{equation}
\frac{d \sigma^0}{d  \Omega_{\tau^-}}=
\frac{\alpha^2 }{16 \ s} \beta (2-\beta^2 sin^2 \theta)
\label{cross0}\end{equation}
where $\alpha$ is the fine structure constant, $s=q^2$ is the
square of the 4-momenta carried by the photon, $\theta$ is 
the angle defined by the electron and $\tau$ directions,
 and 
$\gamma=\frac{\sqrt{s}}{2 m_\tau}$,
$\beta=\sqrt{1-\frac{1}{\gamma^2}}$, 
are
the dilation factor and $\tau$ velocity, respectively.
The second term 
 $\frac{d\sigma^{S}}{d\Omega_{\tau^-}}$,
  involves
spin linear
contributions and is suppressed in our hypothesis.
The last term in (\ref{cross1})
is 
proportional to the product of the spins of both $\tau$'s
and can be written as:
\begin{eqnarray}
\frac{d\sigma^{SS}}{d \Omega_{\tau^-}}  &=& 
 \frac{\alpha^2}{16 s} \beta 
 \left( s_+^x s_-^x C_{xx} + s_+^y s_-^y C_{yy} +\right. \nonumber \\
& & \left. s_+^z s_-^z C_{zz} + (s_+^x s_-^y + s_+^y s_-^x) C_{xy}^+ +
 \right. \nonumber\\ & &
\left. 
(s_+^x s_-^z + s_+^z s_-^x) C_{xz}^+ + 
 \right. \nonumber\\ & &
\left. 
(s_+^y s_-^z +s_+^z s_-^y) C_{yz} +\right.\nonumber\\ & &
 \left. (\mbox{\boldmath $s$}_+ \times {\mbox{\boldmath $s$}}_-)_x
 C_{yz}^- + 
(\mbox{\boldmath $s$}_+ \times \mbox{\boldmath $s$}_-)_y C_{xz}^- + 
\right. \nonumber \\
& & \left. 
(\mbox{\boldmath $s$}_+ \times \mbox{\boldmath $s$}_-)_z C_{xy}^-
\right)
\label{csection}\end{eqnarray}
where
\bea
C_{xx} &=&
(2-\beta^2) \sin^2\theta\nonumber
\\C_{xz}^+ &=& \frac{1}{\gamma} sin 2 \theta
\nonumber
\\
C_{yy} &=& - \sin^2\theta
\nonumber
\\C_{xy}^- &=& 2 \beta \sin^2\theta\, d^\gamma_\tau\nonumber
\\
C_{zz} &=& (\beta^2+(2-\beta^2) \cos^2\theta) 
\nonumber
\\C_{yz}^- &=& \gamma\beta  sin^2\theta\, d^\gamma_\tau
\label{cross3}
\eea


The spin
properties of the
produced taus translates in the angular distribution of both tau decay
products; in order to have access to the EDM one has to measure this
angular distribution. It is  by means of asymmetries that 
this may be done. In this way we select each one of
the terms we are interested in. 
 In what follows we will show how to measure  the
terms that contain the EDM. We will sum in all kinematic
variables as possible in order to enlarge the signal.
 The EDM
is the leading contribution to the normal-transverse ($y-x$) and
normal-longitudinal ($y-z$) correlations. 
We now show how to define an observable proportional to the $C_{xy}^-$ term.
In each of the correlation terms in the cross section
 we have several kinematic variables to take
into account: the CM angle $\theta $ of production of the tau with respect to the
electron, the azimuthal $\phi_{h^+}$, $\phi_{h'^-}$ and polar 
$\theta_{h^+}$, $\theta_{h'^-}$ angles of 
the produced hadrons $h^+$ and $h'^-$ in the $\tau^\pm$
 rest frame.
Both hadron momentum
are fixed by energy conservation and the neutrino in each channel 
will be integrated out.

The  $C_{xy}^-$ term  in the cross section $d\sigma (e^+e^- \rightarrow
\gamma \rightarrow \tau^+\tau^- 
\rightarrow h^+\bar{\nu} h'^-\nu)$ can be written as:

\begin{eqnarray}
\frac{d\sigma^8}{d\Omega_\tau  d^3 q_-^* d^3 q_+^*}  \Biggr|_{C_{xy}^-}
\hspace*{-4mm}&=&
\hspace*{-3mm} 
\frac{\alpha^2\beta^2}{128 \pi^3 s^2} Br_+\, 
Br_- \times\nonumber\\
\hspace*{-4mm}&&d^\gamma_\tau \, \sin^2\theta
\times\nonumber\\
\hspace*{-4mm} 
 & &(n_{+x}^*n_{-y}^* - 
n_{+y}^*n_{-x}^*) \times\nonumber\\ \hspace*{-4mm} 
& &\delta(q_-^*-P_-)\delta(q_+^*-P_+)
\end{eqnarray}
where  $Br_+=Br(\tau^+ \rightarrow h^+\nu)$, 
$Br_-= Br(\tau^- \rightarrow h'^-\overline{\nu})$, $\mbox{\boldmath $n$}_\pm^*= \pm\alpha_\pm 
\hat{\mbox{\boldmath $q$}}_\pm^*$; $\alpha_\pm$ are the polarization
parameters of the $\tau$ decay, $\mbox{\boldmath $q$}_\pm$ are the 
momentum of the hadrons, 
$P_\pm= \frac{m_\tau^2-m_\pm}{2 m\tau}$; all $*$ means that the quantities 
are referred to the $\tau$ rest frame of reference.
Integrating in some of the angles we end up with
\bea
\frac{d^2\sigma}{d\phi_-^* d\phi_+^*} 
&=& \frac{\alpha^2\beta^2}{192 s^2}Br_- Br_+ 
\alpha_-\alpha_+\times\nonumber\\& &  \sin
(\phi_-^* - \phi_+^*) \,d^\gamma_\tau
\eea
We now  integrate in these angles and define the following asymmetry
\begin{equation}
A_{NT} = \frac{
d\sigma_+-d\sigma_-}{d\sigma_++d\sigma_-}
\end{equation}
where
\bea
d\sigma_+&=&\displaystyle{
\int_{\sin(\phi_-^*-\phi_+^*)>0}
\frac{d^2\sigma}{d\phi_-^* d\phi_+^*}
} \,d\phi_-^*\,d\phi_+^* \\
d\sigma_-&=&\displaystyle{
\int_{\sin(\phi_-^*-\phi_+^*)<0}
\frac{d^2\sigma}{d\phi_-^* d\phi_+^*}
}\,d\phi_-^*\,d\phi_+^*
\eea

A straightforward computation gives
\beq
A_{NT} = \frac{4 \beta}{ \pi} \frac{\alpha_-\alpha_+ }{3 - \beta^2}
d^\gamma_\tau \label{antphi+-}\eeq
We have verified that all other terms in the cross section, i.e. the 
spin independent ones, the ones coming with the linear 
polarization and all the other spin-spin correlation terms,  are
eliminated when we integrate in the way we propose.
This means that the only  contribution to this asymmetry is 
exactly the term $C_{xy}^-$ we are interested in. In this way we
have defined a normal-transverse correlation 
observable directly proportional to the EDM.

In a similar way, we can define an observable related 
to the normal-longitudinal correlation term. In this case the angular 
dependence on  the decay product of both $\tau$ is different and there
are various observables we can define. For example,  we integrate in the
regions where  $\sin\theta_-^*\sin\phi_-^*\cos\theta_+^*$ is positive
and negative to obtain
\beq
A^¯_{NL} = - \frac{4}{  \pi^2}\frac{\beta\gamma 
 }{3 - \beta^2} \alpha_- \alpha_+\, d^\gamma_\tau
\label{anlphi+-}
\eeq
The above expressions are  linear in the spin $\alpha$ factors; 
we can  sum all over the channels of both  $\tau$'s in 
order to enlarge the asymmetry.

All the above ideas can be applied for 
$e^+e^-$ collisions at the $\Upsilon$ peak 
where the $\tau$ pair production is  
$e^+e^- \rightarrow \Upsilon \rightarrow \tau^-\tau^-$. 
In this way we may have an important  tau pair production rate.
We are interested in $\tau$ pairs produced by the 
 decays of the $\Upsilon$ resonances, therefore 
we can use $\Upsilon(1S)$, $\Upsilon(2S)$ and $\Upsilon(3S)$, but not 
$\Upsilon(4S)$ because it decays dominantly into $B\overline{B}$. We assume that
the resonant diagrams (b) and (d) of Fig. 1. dominate the process 
on the $\Upsilon $ peak. The
 $\Upsilon$ propagates with a Breit-Wigner 
and the $F_\Upsilon (q^2)$  vector form factor defined as

\beq
\langle \Upsilon(w,\mbox{\boldmath $q$}) | 
\bar{\psi}_b\gamma_\mu\psi_b(0)|0\rangle 
= F_\Upsilon (q^2) \epsilon^*_\mu(w,\mbox{\boldmath $q$})
\eeq
is related to the partial width of $\Upsilon 
 \rightarrow e^+e^-$,

\beq
\Gamma_{ee} = \frac{1}{6\pi}Q_b^2
\frac{(4\pi\alpha)^2}{M_\Upsilon^4} |F_\Upsilon|^2 \frac{M_\Upsilon}{2}
\eeq
where $Q_b = - \frac{1}{3}$ in the electric charge of the $b$ quark.
All the hadronic physics in our process is 
included in this  form factor.


The tau pair production  at the $\Upsilon$ peak introduces the same
polarization
matrix terms with respect  to the production with $\gamma$ exchange
(diagrams (a) and (b)). The only difference is 
 an overall factor $
\left(\frac{e^2 Q_b^2  |F_\Upsilon|^2}{s \Gamma_\Upsilon
M_\Upsilon}\right)^2 = 
\left(\frac{3}{\alpha} Br(\Upsilon \rightarrow e^+e^-)\right)^2$
that is introduced in the cross section  ($s = M_\Upsilon^2$).
The only contributions with  EDM in the polarization terms  
come  with the interference of diagrams (b) and (d) while 
diagram (b) squared gives the tree level terms. All the comment we did
with respect to Eqs.(\ref{cross1}), (\ref{cross0}) and 
(\ref{csection}) are useful here, and  we find that there are no changes
in the asymmetries
 and their expression  at the 
$\Upsilon$ peak remain the same as before.

For $10^7 - 10^8$ $\tau$'s the asymmetries provide  bounds of the order
of $10^{-17} - 10^{-18} \,e\,cm$ for the EDM. This is   one or two
orders of magnitude lower than the last PDG bound. To  finish we would
like to stress that:

1) with low energies data  we may ahve  an independent analysis of the EDM from
that obtained with LEP data,

2) low energies data makes possible a clear separation of the effects
coming from
the EDM and the weak-EDM,

3) high statistics can  compensate the suppression factor
$q^2/\Lambda^2$ in the low energy regime for the effective operators,

4) as contrary as it is  now the case in the published bounds  
 our observables are {\em CP-odd} ones and the limits thus 
obtained are rigorous.


I would like to thank the Organizers   for a very pleasant and
interesting week during the Tau-2002  Workshop. This work has been
supported by CONICYT-Uruguay, CICYT-Spain BFM2002-00568 and 
FPA2002-00612 and 
by Agencia Espa\~nola de Cooperaci\'on Internacional. 

\end{document}